\documentclass{article}
\usepackage{graphicx}           % for eps figures
\usepackage{bm}                 % for bold math symbols
\usepackage{amsmath}            % for \text and such
\usepackage{amssymb}            % more symbols...
\usepackage{verbatim}           % useful for commenting out stuff
\usepackage{amsthm}             % does  gs etc. nicely, but
\theoremstyle{plain}            % This is the default
\newtheorem{theorem}{Theorem}

\newtheorem{lemma}{Lemma}

\newcommand{\arcsinh}{{\rm arcsinh}}
\newcommand{\arccosh}{{\rm arccosh}}

\def\bbfZ{{\hbox{Z}\!\!\hbox{Z}}}

\title{Evanescence in Coined Quantum Walks}
\author{Hilary A. Carteret \\
Laboratoire d'Informatique Th\'eorique et Quantique, \\
D\'epartement d'Informatique et de Recherche Op\'erationelle, \\
Pavillon Andr\'e-Aisenstadt, \\
             Universit\'e de Montr\'eal, \\ 
Montr\'eal, Qu\'ebec, H3C 3J7 \\ 
Canada \\
email: cartereh@iro.umontreal.ca
\and Bruce Richmond \\
Department of Combinatorics and Optimization \\
             University of Waterloo \\ 
Waterloo, Ontario, N2L 3G1 \\ 
Canada \\
email: lbrichmo@hopper.math.uwaterloo.ca
\and Nico M. Temme \\
CWI \\
Kruislaan 413, NL-1098, SJ Amsterdam, \\
Netherlands, \\ 
email: Nico.Temme@cwi.nl}
\date{$10^{{\rm{th}}}$ August, 2005}

\begin{document}

%%%%%%%%%%%%%%%%%%%%%%%%%%%%%%%%%%%%%%%
%% Actually produce the title first. 
%% Silence this with a % to save paper
\maketitle

\begin{abstract}
In this paper we complete the analysis begun by two of the authors in
a previous work on the discrete quantum walk on the infinite line 
[J. Phys. A 36:8775-8795 (2003); quant-ph/0303105].
We obtain uniformly convergent asymptotics for the ``exponential decay'' 
regions at the leading edges of the main peaks in the Schr{\"o}dinger 
(or wave-mechanics) picture.  
This calculation required us to generalise the method of stationary 
phase and we describe this extension in some detail, including 
self-contained proofs of all the technical lemmas required.  
We also rigorously establish the exact Feynman equivalence between the 
path-integral and wave-mechanics representations for this system using 
some techniques from the theory of special functions.  Taken together 
with the previous work, we can now prove every theorem by both routes.  
\end{abstract}

\tableofcontents

%--------------------- end of title stuff --------------------------%

%%%%%%%%%%%%%%%%%%%%%% Now begin the text %%%%%%%%%%%%%%%%%%%%%%%%%%%

\section{Introduction}

The first authors to discuss the quantum walk were Aharonov, Davidovich 
and Zagury in \cite{earliest} where they described a very simple realization 
in quantum optics. In this model a particle takes unit steps on the integers 
at each time step, starting at the origin.  
In \cite{Meyer96} Meyer proved that an additional spin-like degree of 
freedom was essential if the behaviour of the system was to be both unitary 
and non-trivial.  Without this degree of freedom, the only way the evolution 
of the walk can avoid being purely ballistic is to relax the unitarity 
condition. This spin-like degree of freedom is sometimes called the 
{\emph{chirality,}} or the {\emph{coin}}, which is why this type of walk 
is sometimes called a ``coined'' walk \cite{Watrouswalk}.  
This is in sharp contrast to the continuous time walk \cite{cqwalk1,cqwalk2} 
which does not need a coin and which we will not discuss here.
The chirality can take the values {\small{RIGHT}} and {\small{LEFT}}, or a 
coherent superposition of these.  
\begin{figure}[floatfix]
  \begin{minipage}{\columnwidth}
   \begin{center}
    \resizebox{0.8\columnwidth}{!}{\includegraphics{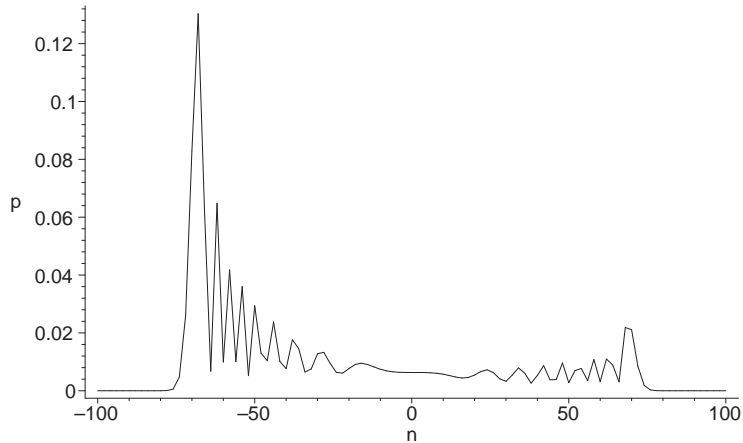}}
   \end{center}
  \end{minipage}
  \caption{{\small The discrete quantum walk on the line.  The probability 
           distribution is shown for a walk that started at the origin
           with its coin in the state $|\rm{L}\rangle$ after it has evolved 
           for 100 steps.  The distribution is oscillatory between the two 
           peaks and decays exponentially outside that range.  The peaks 
           move away from the origin with speed $t/\sqrt{2},$ and the width 
           of the peaks also decreases with time.  By contrast, the classical 
           random walk has a Gaussian distribution, which spreads with 
           velocity $\sim \sqrt{t}.$}} \label{walk}
\end{figure}
For a detailed introduction to quantum walks, we refer the reader to the 
review article in \cite{JuliaReview}.

Meyer and subsequent authors \cite{Aharonov00,Ambainis01} have considered 
two approaches to the discrete-time quantum walk, the {\bf{path-integral}} 
approach of Feynman and the Schr\"{o}dinger {\bf{wave-mechanics}} approach, 
which reflect two complementary ways of formulating quantum mechanics 
\cite{FeynmanHibbs}. 
We refer to the paper by Ambainis, Bach, Nayak, Vishwanath and Watrous 
\cite{Ambainis01} for proper definitions and more references.  Both 
approaches are discussed in the paper of Ambainis {\emph{et al.}} The 
probability distribution for this walk is shown in Figure~\ref{walk}, after 
enough time has elapsed for the asymptotic behaviour to manifest. 

\smallskip

This paper began as a sequel to the work of Carteret, Ismail and Richmond  
\cite{3routes} concerning the one-dimensional quantum walk on the integers, 
and contains the completion of the analysis of this quantum walk for the 
remaining exponential decay region in the Schr{\"{o}}dinger picture.  We 
believe the analysis presented in this paper to be interesting for three 
reasons.  One is methodological; while analysing this system we encountered 
various links between a number of different methods in combinatorics which 
do not seem to be widely known, and which may be of use to the quantum 
information community when analysing more complicated systems than the 
one discussed here. 

The second motivation is rather more abstract.  It is one of the fundamental 
principles of quantum mechanics that the wave-mechanics and path-integral 
representations of a system should produce {\emph{exactly the same results.}} 
The quantum walk has been proposed as the quantum analogue of the classical 
random walk, in the hopes of ultimately defining a systematic procedure for 
``quantizing'' classical random walk algorithms 
\cite{JuliaReview,Ambwalk,Szegedy}. 
Quantizing classical systems is something that must be done with 
considerable care; the obvious approach isn't necessarily the correct one.  
While it is true that when such pathologies have been discovered in the 
past, they were found in much more exotic systems than this one, it is as 
well to check.  One way to perform such a check is to verify that the 
Feynman equivalence principle still holds between the path-integral and 
wave-mechanics representations.  
As it happens, the results obtained from the two approaches do not at 
first appear to be equivalent. In fact they are, though the proof of 
their equivalence is nontrivial, and will be given in 
Section~\ref{equivalence}.  In the course of this analysis we 
uncovered a small, but potentially significant omission in previous 
analyses of this system, which we will describe below.  

Another reason for performing this check is to explore the little mystery 
left at the end of \cite{3routes}. While the results from the path integral 
calculation made intuitive sense, the partial results from the wave-mechanics 
calculation for the exponential decay region were rather unexpected.  
Specifically, we found what appeared to be evanescent waves in this 
exponential decay region, which seemed to imply the presence of some kind 
of absorption mechanism, despite the fact that the definition of the model 
precludes any barrier or other source of dissipation in the system to 
cause these by the familiar mechanisms.  So, we will check that the Feynman 
equivalence holds to verify that those evanescent waves are not simply some 
kind of mathematical artefact.

We would also like to gain some insight into the physical interpretation 
of the mathematical behaviour of this part of the wave-function.  A link 
between the behaviour of the quantum walk on the line and certain phenomena 
in quantum optics has been suggested previously by Knight, Roldan and Sipe in 
\cite{origininf,opticimpl,interfwalk}, and refined by Kendon and Sanders 
in \cite{complementarity}.  This connection will be discussed in more 
detail in Section~\ref{interpretation}.  

We will also describe a potentially useful method for obtaining integral 
representations of orthogonal polynomials from their generating functions 
using Lagrange Inversion.  This bypasses the need to use the Darboux 
method and makes it possible to obtain uniformly convergent asymptotics 
directly from the generating function.  We have included this in the 
Appendices in Subsection~\ref{lagrange}.

\subsection{Some previous results}\label{previous}

In this section we will mention some results by other authors which we will 
have occasion to use later in this paper.  One of the reasons for doing this 
is that different authors have used different labelling conventions and this 
will enable us to establish a consistent notation for use when we combine 
results from different papers with mutually incompatible conventions.  We 
will state our results using the conventions in \cite{3routes}.

Several early results in the theory of quantum walks are due to Meyer 
\cite{Meyer96}, who considered the wavefunction as a two-component vector 
of amplitudes of the particle being at point $n$ at time $t$. Let
\begin{equation}
\Psi(n,t)=
  \begin{pmatrix}
  \psi_{R}(n,t) \\
  \psi_{L}(n,t)
 \end{pmatrix}
\end{equation}
where the chirality of the top component is labelled {\small{RIGHT}} and the 
bottom {\small{LEFT}}. At each step the chirality of the particle evolves 
according to a unitary Hadamard transformation
\begin{align}
|R\rangle &\mapsto \frac{1}{\sqrt{2}}\left( |R\rangle + |L\rangle\right) \\
|L\rangle &\mapsto \frac{1}{\sqrt{2}}\left( |R\rangle - |L\rangle\right),
\end{align}
which is why this quantum walk is sometimes called the ``Hadamard'' walk.
The particle (or ``walker'') then moves according to its new chirality 
state. Therefore, the particle obeys the recursion relations
\begin{align}
 \Psi_{R}(n,t+1) &=  \frac{1}{\sqrt{2}}\Psi_{L}(n+1,t)
   +\frac{1}{\sqrt{2}}\Psi_{R}(n-1,t) \\
 \Psi_{L}(n, t+1) &=  -\frac{1}{\sqrt{2}}\Psi_{L}(n+1,t)
   +\frac{1}{\sqrt{2}}\Psi_{R}(n-1,t).
\end{align}

\bigskip

Meyer approached this problem from the {\bf{path-integral}} point of view, 
and obtained expressions for the $\psi$-functions in terms of Jacobi 
polynomials. The standard notation in \cite{AnStegun} for Jacobi polynomials 
is $P_{n}^{(\alpha,\beta)}(z)$ but we wish to follow the conventions in 
\cite{NayaknV} and subsequent papers that have now become standard in the 
literature on quantum walks, and use $\alpha=n/t.$ 
We will therefore use the notation $J_{q}^{(r,s)}(w).$ 

\noindent
We find, in \cite{Ambainis01} and \cite{3routes} when $n \ne t$,
\begin{theorem}[Ambainis {\emph{et al.}} \cite{Ambainis01}, after Meyer, 
\cite{Meyer96}]\label{Amb1}
\begin{equation}
\psi_{R}(n,t) (-1)^{(t-n)/2}=
\begin{cases}
 2^{n/2-1}J_{((t+n)/2-1)}^{(1,-n)}(0) &\text{when }-t \le n <0 \\
\left( \frac{t+n}{t-n}\right)2^{-n/2-1}J_{(t-n)/2-1}^{(1,n)}(0)
 &\text{when } 0\le n <t
\end{cases}
\end{equation}
Also
\begin{equation}
\psi_{L}(n,t)(-1)^{(t-n)/2}=
\begin{cases}
2^{n/2}J_{(t+n)/2}^{(0,-n-1)}(0) &\text{when }-t \le n <0 \\
2^{-n/2-1}J_{(t-n)/2-1}^{(0,n+1)}(0) &\text{when }0 \le n <t
\end{cases}
\end{equation}
and
\begin{align}
  |\psi_{R}(n,t)|^{2} 
  &=\left(\frac{t-n}{t+n}\right)^{2}|\psi_{R}(-n,t)|^{2}\label{rightAmb1} \\
  |\psi_{L}(n,t)|^{2} &=|\psi_{L}(2-n,t)|^{2}\label{leftAmb1}.
\end{align}
\end{theorem}

\smallskip

Ambainis {\emph{et al.}} use the other sign convention, so one should 
interchange $L$ and $R$ (or equivalently, replace $n$ by $-n$) to reflect 
the walk before comparing their results with ours.  This is just a 
relabelling, and so their results can be stated as in the following theorem.
We will prove the above results in the form below
\begin{theorem}[Ambainis {\emph{et al.}} \cite{Ambainis01}]\label{Amb2} 
When $ n \equiv t\,({\rm{mod}} \, 2)$ and $J_{q}^{(r,s)}(w)$ denotes a Jacobi 
polynomial, then
\begin{equation}\label{psiRJac}
\psi_{R}(n,t)(-1)^{(t-n)/2}=
\begin{cases}
(-1)^{n+1}2^{-n/2}J_{(t-n)/2}^{(0,n-1)}(0) &\text{when } 0 \le n \le t \\
(-1)^{n+1}2^{n/2 -1} J_{(t+n)/2-1}^{(0,-n+1)}(0)   &\text{when } -t < n <0
\end{cases}
\end{equation}
Also
\begin{equation}\label{psiLJac}
\psi_{L}(n,t)(-1)^{(t-n)/2}=
\begin{cases}
  -2^{-n/2-1}J_{(t-n)/2 -1}^{(1,n)}(0) &\text{when }0\le n <t \\
 -\left(\frac{t-n}{t+n}\right)2^{n/2-1}J_{(t+n)/2-1}^{(1,-n)}(0) 
  &\text{when } -t< n <0.
\end{cases}
\end{equation}
and
\begin{align}
 \psi_{R}(-n,t) &= (-1)^{n+1}\psi_{R}(n+2,t)\label{leftAmb2}, \\ 
 (t-n)\psi_{L}(-n,t) &= (-1)^{n}(t+n) \psi_{L}(n,t)\label{rightAmb2}.
\end{align}
\end{theorem}

\bigskip
\noindent
{\bf{A few Remarks:}} 
\begin{enumerate}
\item

Note that Theorem~\ref{Amb2} differs from Theorem~\ref{Amb1} by an external
phase which has been dropped in previous analyses of this system; we state 
the symmetry relations for the $\psi$-functions rather than for their 
moduli-squared, as in \cite{Ambainis01}.  We will discuss this in more 
detail below, as some properties of Jacobi polynomials are required for 
the calculation. 

\item
There is a sign error in the symmetry relations for $\psi_{R}$ and 
$\psi_{L}(-n,t)$ in Carteret {\emph{et al.}} \cite{3routes} (which has 
been corrected in the arxiv version \cite{typo}). The symmetry relations 
will be proved in Lemmas~\ref{lempsir}, \ref{lempsil} and equation 
\eqref{jacsym} of Subsection~\ref{symmetries} using some integral 
representations of $\psi_{R}(n,t)$ and $\psi_{L}(n,t)$.

\item
The endpoints where $n=\pm t$ have to be handled separately, 
see \cite{Meyer96}. For the starting conditions 
\begin{align}
  &\psi_{L}(0,0) = 1, \nonumber
  &\psi_{R}(0,0) = 0  \nonumber
\end{align}
the wavefunctions at the end-points (where $n = \pm t$) are 
\begin{align}\label{endpoints}
  \psi_{R}(t,t) &= (-1)^{t+1}2^{-t/2} &t=0,1,2,\ldots, \\
  \psi_{L}(t,t) &= 0, &t=1,2,3,\ldots,\\ 
  \psi_{R}(-t,t)&= 0, &t=1,2,3,\ldots,\\ 
  \psi_{L}(-t,t)&= (-1)^t2^{-t/2}, &t=0,1,2,\ldots.
\end{align} 

\item
The two different cases in \eqref{psiLJac} and \eqref{psiRJac} for $n\ge 0$
and $n<0$ can be combined into one case for all $n$ satisfying
$-t\le n<t.$ We prove this later using a symmetry property of the Jacobi 
polynomials.  Our results in equations \eqref{leftAmb2} and \eqref{rightAmb2}
are refinements of the corresponding relations in \eqref{leftAmb1} and 
\eqref{rightAmb1}, after performing the relabelling necessary to compare 
results with different sign conventions.

\end{enumerate}

The asymptotic behaviour for the path-integral representation has been 
determined in Carteret {\emph{et al.}} \cite{3routes}, starting from 
Theorem~\ref{Amb2}.  The steepest descent technique was used on the 
standard integral representation for the Jacobi polynomial.  The result was 
uniform exact asymptotics $\alpha$ in the range $|\alpha|<1-\varepsilon,$ 
where $\varepsilon$ is any positive number, in terms of Airy functions.

This technique was used earlier for Jacobi polynomials by Saff and Varga
\cite{Saff} and by Gawronkski and Sawyer \cite{Gawyer}; however the 
connection with Airy functions had not been recognized as far as we know. 
The Airy function description is useful for $|\alpha |$ near $1/\sqrt{2}$ 
where the asymptotic behaviour changes from an oscillating cosine term times 
$t^{-1/2}$ (for $|\alpha|<1/\sqrt{2}$) to exponentially small 
(for $2^{-1/2}+\varepsilon < |\alpha| <1-\varepsilon).$

\bigskip

In this paper we analyze the Hadamard walk from the Schr\"odinger 
{\bf{wave-mechanics}} point of view.  The earliest work on this that the 
authors are aware of is that by Nayak and Vishwanath \cite{NayaknV}. 
They define 
\begin{equation}
 \widetilde{\Psi}(\theta,t)=\sum_{n}\psi(n,t)e^{i\theta n}, 
\end{equation}
(where we have used the symbol $\theta$ for the momentum instead of the $k$ 
used in \cite{NayaknV}) so the recursion relations above becomes
\begin{equation}
 \widetilde{\Psi}(\theta,t+1)=M_{\theta} \widetilde{\Psi}(\theta,t),
\end{equation}
where 
\begin{equation}
M_{\theta}=
\begin{pmatrix}
 e^{-i\theta} & e^{-i \theta} \\
 e^{i \theta} & -e^{i\theta}
 \end{pmatrix}\,.
\end{equation}
Thus
\begin{equation}
 \widetilde{\Psi}(\theta, t)=M_{\theta}^{t}\widetilde{\Psi}(\theta,0), \quad
\widetilde{\Psi}(\theta,0)=(1,0)^{{\rm{T}}},
\end{equation}
where the symbol $\;^{{\rm{T}}}$ denotes transposition. 
They show that the eigenvalues of $M_{\theta}$ are $e^{-i\omega_{\theta}}$ 
and $-e^{i\omega_{\theta}}$ where $\omega_{\theta}$ is the 
angle in $[-\pi/2, \pi/2]$ such that $\sin(\omega_{\theta})=
(\sin \theta)/\sqrt{2}.$ They also use the other sign convention, so one 
should relabel $L$ and $R$ as before.  Their results can be stated as in 
the following theorem.
\begin{theorem}[Nayak and Vishwanath \cite{NayaknV}]\label{NV1}
Let $\alpha = n/t$. Then
\begin{equation}\label{psiRintegral}
  \psi_{R}(n,t)=\frac{1+(-1)^{n+t}}{2} \frac{1}{2\pi} 
  \int_{-\pi}^{\pi} \frac{e^{i \theta}}{\sqrt{1+\cos^{2} 
  \theta}}e^{-i(\omega_{ \theta}+\theta \alpha )t}\; d \theta.
\end{equation}
\begin{equation} \label{psiLintegral}
 \psi_{L}(n,t)=\frac{1+(-1)^{n+t}}{2} \frac{1}{2\pi} 
 \int_{-\pi}^{\pi}\left(1+\frac{\cos 
 \theta}{\sqrt{1+\cos^{2}\theta}}\right)
 e^{-i(\omega_{\theta}+\theta \alpha )t}\; d \theta, 
\end{equation}
\end{theorem}

We will derive the asymptotic behaviour of the $\psi$-functions starting 
from Theorem~\ref{NV1} in Section~\ref{genstat}.  We will only give the 
complete details for the exponential decay range, as the calculation for 
the oscillatory region has already been done by others 
\cite{NayaknV,Ambainis01}.  The conventional version of the method of 
stationary phase, as used by Nayak-Vishwanath \cite{NayaknV}, does not work 
in the exponentially small region, as the stationary points of the phase 
function have left the real line.  We will show how to extend and refine 
this technique so that it can be made to work in this situation; the 
modification is an application of the method of steepest descents. 

It is not obvious that the formul\ae\, obtained by each method are the same, 
but we will prove below in Section~\ref{equivalence} that they are. This 
means that precisely the same asymptotic behaviour can be found using both 
the path-integral and wave-mechanics descriptions of quantum mechanics.

\section{Generalizing the method of stationary phase}\label{genstat}

\begin{figure}[floatfix]
  \begin{minipage}{\columnwidth}
   \begin{center}
    \resizebox{0.8\columnwidth}{!}{\includegraphics{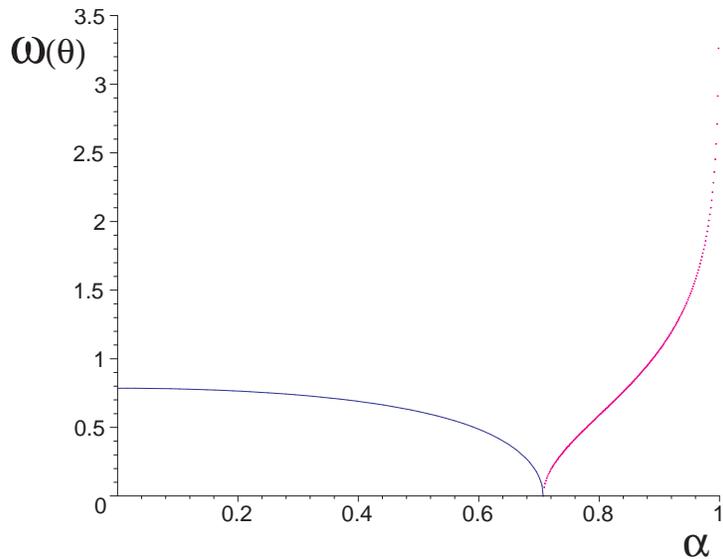}}
   \end{center}
  \end{minipage}
  \caption{{\small The behaviour of $\omega_{\theta}$ as $|\alpha| > 
           1/\sqrt{2}.$  As $|\alpha|$ moves beyond the critical value of 
           $1/\sqrt{2}$, the phase function $\omega_{\theta}$ becomes 
           imaginary.  The behaviour of the real part is shown in the solid 
           line and that of the nascent imaginary part is shown in the 
           dotted line. Multiplying this imaginary phase function by $i$ 
           gives a simple exponential decay.}}\label{omegatrack}
\end{figure}

The aim of this section is to extend the method of stationary phase so that 
it can cope with stationary points that occur as complex conjugate pairs on 
either side of the real axis.  We start with the integral representation of 
$\psi_{R}$ in Theorem~\ref{NV1}.  In this representation the integration is 
performed along the real axis.  Nayak and Vishwanath consider the case 
$|\alpha| < 1/\sqrt{2}$ when there are two stationary points (defined below) 
inside the interval of integration $[-\pi,\pi]$.  When we find the critical 
points of the phase function, we obtain an equation for $\theta$ (called $k$ 
in \cite{3routes}) at the critical points as a function of $\alpha,$ which is 
\begin{equation}\label{kcrit}
 \cos \theta_{\alpha} = \frac{-\alpha}{\sqrt{1-\alpha^2}}
\end{equation}
from which the critical value of $\omega_{\theta},$ call it 
$\omega_{{\theta}_{\alpha}},$ can be obtained using the Pythagoras rule 
($\cos^2 \theta + \sin^2 \theta = 1$) and 
$\omega_{\theta} = \arcsin \frac{\sin \theta}{\sqrt{2}}$ from \cite{NayaknV}.
However, when $|\alpha| > 1/\sqrt{2}$ this equation no longer has any 
{\emph{real}} solutions and thus the corresponding stationary points are 
no longer on the real axis, see figure~\ref{omegatrack}.  Therefore the 
standard method of stationary phase cannot provide the exact asymptotics.   
The stationary points have ``moved'' off the real axis and become a complex 
conjugate pair.  We shall move the contour of integration to follow the 
stationary points, whilst ensuring that the contour still goes through one 
of them.  Note also that the stationary points become saddle-points on 
leaving the real line; we will return to this fact in Subsection 2.2, below.

\subsection{The function ${\boldsymbol{\omega_\theta}}$ in the complex plane}

The key to evaluating the asymptotics for these integrals lies in 
the behaviour of the phase function $\omega_{\theta}.$  We will 
therefore begin by describing the analytic properties of the function
$\omega_{\theta}$ in the complex plane, in particular in the strip
$-\pi\le \Re\theta\le \pi$.  We will need this information when we replace 
the initial interval of integration by a contour in the complex plane, as 
explained in the next subsection.  
We will also need  estimates of $\omega_{\theta}$ at $+ \infty$ in this 
strip to  show that our new contour integral converges.
The singular points of 
$\omega_{\theta}=\arcsin\left(\frac{\sin \theta}{\sqrt{2}}\right)$ are found 
from the equations $\frac{\sin \theta}{\sqrt{2}}=\pm 1$. When we write 
$\theta=u+iv$, with $u\in [-\pi,\pi]$ and $v\in \mathbb{R},$ we conclude 
from the equations $\sin(u+iv)=\pm\sqrt{2},$ where
\begin{equation}\label{sinuv}
    \sin(u+iv)= \sin u\cosh v+i\cos u\sinh v
\end{equation}
that $u=\pm\frac12\pi$ and $\cosh v=\sqrt2$ (or $v=\pm\arcsinh(1)$).
Because of conjugation and symmetry there are four singular points, 
namely $\pm\frac12\pi +i\arcsinh(1)$ and $\pm\frac12\pi -i\arcsinh(1)$.
On the four half lines $\pm\frac12\pi +i v$ with $|v| > \arcsinh(1)$ the 
function $\sin(u+iv)$ is real, and has an absolute value greater than 
$\sqrt2$.
These four half-lines are taken as branch cuts of the multi-valued 
function $\omega_{\theta}$. They correspond with the two branch cuts
of the function $\arcsin z=\arcsin (x+iy)$ in the $z-$plane from 
$x=\pm1$ to $x=\pm\infty$, with $y=0$.

\begin{figure}[floatfix] 
  \begin{minipage}{\columnwidth}
   \begin{center}
    \resizebox{1.0\columnwidth}{!}{\includegraphics{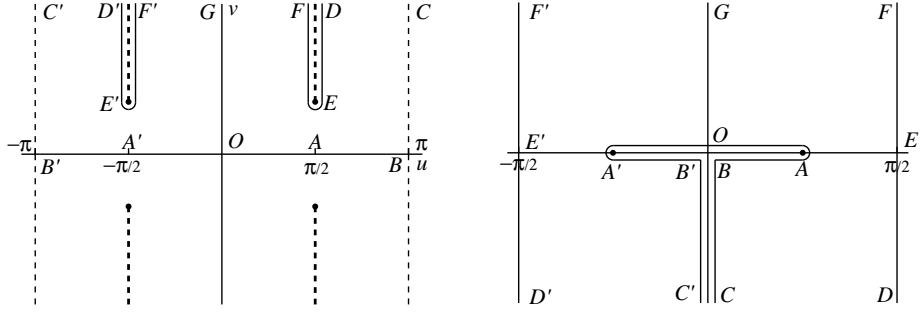}}
   \end{center}
  \end{minipage}
  \caption{{\small Conformal mapping 
          $\omega_\theta=\arcsin\left(\frac{\sin \theta}{\sqrt{2}}\right)$.
          At the left is the $\theta-$plane, $\theta=u+iv$, 
          $-\pi\le u\le \pi$, with the four branch cuts.  At the right is 
          the $\omega_\theta-$plane; only the image of the half-strip
          $-\pi\le u\le \pi$, $v\ge 0$ is shown. }}\label{conformal}
\end{figure}

The strip $-\pi\le u\le \pi$ that is delineated by these branch cuts is 
the principal Riemann sheet on which $\omega_{\theta}$ is analytic and 
single-valued. We consider the principal branch of this function that is 
real on $[-\pi,\pi]$ and continuously extended on the principal sheet. 
Since the function is periodic, the same holds for the other strips 
$[k\pi,(k+2)\pi]$,  $k \in \bbfZ$.

In Figure~\ref{conformal} we show the conformal mapping by $\omega_\theta$ 
from the strip $-\pi\le u\le\pi$.
We show the images of a number of lines, where we concentrate on $v\ge0$. 
For $v\le0$ a similar picture can be given.
We observe the following useful facts.
\begin{enumerate}
\item
The image of the interval $[0,\pi]$ must go around a branch cut because 
$\omega_\theta$ is not single valued on this interval; the image point 
$A$ is given by $A=\arcsin\frac1{\sqrt2}$.

\item
The points $D$ and $F$ 
are on different sides of the branch cut; the loop $DEF$ around the branch 
cut is mapped to the vertical $DEF,$ and the same goes for the three points 
$D'E'F'$. 

\item
On the positive imaginary axis $u=0, v\ge 0$  $\omega_\theta$ has the form
(cf. \eqref{sinuv})
\begin{equation}\label{og}
\omega_\theta =i\arcsinh\frac {\sinh v}{\sqrt2}=
i\ln\left(\frac{\sinh v}{\sqrt2}+\sqrt{\frac{\sinh^2v}{2}+1}\right),
\quad  v \ge 0.
\end{equation}

\item
On the half-lines $u=\pm \pi, v\ge 0, \; \omega_{\theta}$ has the form
\begin{equation}\label{bc}
 \omega_\theta =-i\arcsinh\frac {\sinh v}{\sqrt2}=
 -i\ln\left(\frac{\sinh v}{\sqrt2}+\sqrt{\frac{\sinh^2v}{2}+1}\right),
 \quad  v \ge 0.
\end{equation}

\item
For $\theta$ on the vertical $ED'$ we can write $\omega_\theta$ in the form
\begin{equation}\label{edp}
 \omega_\theta =\frac12\pi+i\ln\left(\frac{\cosh v}{\sqrt2}+
 \sqrt{\frac{\cosh^2v}{2}-1}\right),\quad \cosh v \ge \sqrt2.
\end{equation}

\item
For $\theta$ on $ED$ we must choose the negative square root, which gives
\begin{equation}\label{ed}
 \omega_\theta =\frac12\pi-i\ln\left(\frac{\cosh v}{\sqrt2}+
 \sqrt{\frac{\cosh^2v}{2}-1}\right),\quad \cosh v \ge \sqrt2.
\end{equation}
\end{enumerate}

\noindent
We will also need the following lemma, in order to bound $\omega_{\theta}$  
as the imaginary part of $\theta$ tends to $+\infty,$ and hence guarantee 
that these integrals will converge.

\begin{lemma}\label{bound1}
If $\theta=u+iv,v >0$, then, as $v\rightarrow+\infty$
\begin{equation}\label{fls}
      e^{-i\omega_{\theta}t}=
      \begin{cases}
         \mathcal{O}(e^{+ v t}), & |u|<\frac12\pi,\\
      
         \mathcal{O}(e^{- v t}), & \frac12\pi < |u|\le \pi.
      \end{cases}
\end{equation}
\end{lemma}
\noindent
\begin{proof} Since 
$\omega_{\theta}=\arcsin\left(\frac{\sin \theta}{\sqrt{2}}\right)$ 
we have $\sin \omega_{\theta}=\sin 
\theta/\sqrt{2}$; so
\begin{equation}
  e^{-i\omega_{\theta}}=\cos\omega_{\theta}-i\frac{\sin \theta}{\sqrt{2}}.
\end{equation}
Hence
\begin{equation} \label{eiom}
  e^{-i\omega_{\theta}}=\pm\sqrt{1-\frac{\sin^2 \theta}{{2}}}-
  i\frac{\sin \theta}{\sqrt{2}} 
\end{equation}
where the $\pm$ sign in front of the first term has yet to be determined.
For small values of $\theta$, it is obvious that we should select the $+$ 
sign, because both sides of equation \eqref{eiom} have to approach unity as 
$\theta\to 0$. In fact, the $+$ sign should be chosen throughout the strip 
$|u|<\frac12\pi$.  This is because
\begin{equation}
  1-\frac{\sin^2\theta}{2}=\frac{3+\cos2\theta}{4}\sim\frac18e^{-2i\theta}
\end{equation}
as $\theta\to+i\infty$, we conclude that in the strip $|u|<\frac12\pi$
\begin{equation}
  e^{-i\omega_{\theta}}\sim 
  \frac1{2\sqrt2}e^{-i\theta}-i\frac{\sin\theta}{\sqrt2}
  \sim \frac1{\sqrt2}e^{-i\theta}={{\cal O}}(e^v),
\end{equation}
as $v\to+\infty$. Observe that this is in agreement with the behaviour of
$\omega_{\theta}$ on the positive imaginary axis, as given in equation 
\eqref{og}. It also agrees with the conformal mapping
shown in Figure~\ref{conformal}, where we see that the domain $AEFGF'E'A'OA$
is mapped to $\Im\omega_\theta>0$. The figure also shows that the domains 
$ABCDEA$ and $A'B'C'D'E'A'$ are mapped to $\Im\omega_\theta<0$. This 
corresponds to choosing the negative values for $\theta$ in \eqref{eiom} in 
these domains; this gives the estimate in the second line of \eqref{fls}.
This proves the lemma. 
\end{proof}

\noindent
Now that we have established the behaviour of the phase function 
$\omega_{\theta},$ we can proceed to choose an appropriate contour of 
integration.

\subsection{Saddle-point analysis}\label{saddlepoint}

We will now obtain an asymptotic approximation for $\psi_R$ in the 
exponentially small range outside the main peaks. To find convenient 
locations for the contours of integration with respect to the stationary 
points, we will begin with the integral in \eqref{psiRintegral} of 
Theorem~\ref{NV1} for $\psi_R(2-n,t)$, and use the symmetry rule for 
$\psi_{R}(n,t)$ (cf. Theorem~\ref{Amb2}) to obtain the result for 
$\psi_R(n,t)$. So, our starting point is 
\begin{equation}\label{psiRint}
 \psi_{R}(n,t)= (-1)^{n+1}\psi_{R}(2-n,t)=\frac{(-1)^{n+1}}{2\pi} 
 \int_{-\pi}^{\pi} \frac{e^{-i \theta}}{\sqrt{1+\cos^{2} 
 \theta}}e^{-i(\omega_{ \theta}-\theta \alpha )t}\; d \theta,
\end{equation}
where $\alpha = n/t$. We have dropped the factor $\frac{1+(-1)^{n+t}}{2}$, 
because we always can assume that $n$ and $t$ have the same parity; 
the wavefunction is identically zero otherwise as the walker must always 
move at each time-step.

\bigskip

\begin{figure}[floatfix]
  \begin{minipage}{\columnwidth}
   \begin{center}
    \resizebox{0.6\columnwidth}{!}{\includegraphics{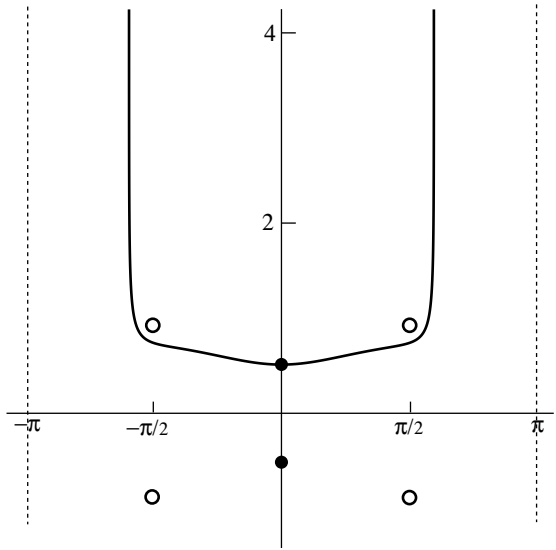}}
   \end{center}
  \end{minipage}
  \caption{{\small{The saddle-point contour for the integral in 
            \eqref{psiRint} for the exponential-decay range, where 
            $1>\alpha > 1/\sqrt{2}$ and the countour runs from $-\pi$ to 
            $-\pi+i\infty.$ It continues from that point through the 
            saddle-point at the positive imaginary axis to $+\pi+i\infty$ 
            and from that point to $+\pi$.  Again, the contributions from 
            the vertical half-lines cancel each other.  On the contour 
            shown, the imaginary part of the phase function 
            $-i(\omega_{\theta}-\theta \alpha)t$ is constant (equal to $0,$ 
            in fact). The real part tends to $-\infty$ in the valleys at 
            infinity, and has a maximum at the saddle-point (black dot on 
            the positive imaginary axis).}}} \label{expodecint}
\end{figure}

We first locate the stationary points or saddle-points in the traditional 
way, that is, we solve the equation
\begin{equation}\label{sad}
\alpha= \frac{d\omega_{\theta}}{d\theta}=\frac{(\cos 
\theta_{\alpha})/\sqrt{2}}{\sqrt{1-\sin^{2}\theta_{\alpha}/2}}=
\frac{1}{\sqrt{1+\cos^{2}\theta_{\alpha}}} 
\cos\theta_{\alpha}.
\end{equation}
Note that in \eqref{sad} $\cos \theta_{\alpha}$ and $\alpha$ have the same 
sign (and $\alpha$ is positive). This gives
\begin{equation}
  \cos\theta_{\alpha}=\pm \frac{\alpha}{\sqrt{1-\alpha^{2}}}.
\end{equation}
Thus, when $\alpha < 1/\sqrt{2},$ this gives two real stationary points
\begin{equation}
\theta_{\alpha}= \pm \arccos\left(\frac{\alpha}{\sqrt{1-\alpha^{2}}}\right),
\end{equation}
which are used in the stationary phase method in \cite{NayaknV}.
If $1/\sqrt{2}<\alpha <1 $ these points are purely imaginary, and they are 
given by
\begin{equation}
 \theta_{\alpha} =\pm i\mbox{arccosh}
\left(\frac{\alpha}{\sqrt{1-\alpha^{2}}}\right).
\end{equation}
When $1/\sqrt{2}<|\alpha| <1 $ we shift the contour in the integral 
representation of the $\psi_R$ given in \eqref{psiRint} off the real axis 
to the segments shown in Figure~\ref{expodecint}.  Our modified stationary 
phase method is in fact a version of the method of steepest descents. 

\begin{figure}[floatfix]
  \begin{minipage}{\columnwidth}
   \begin{center}
    \resizebox{0.6\columnwidth}{!}{\includegraphics{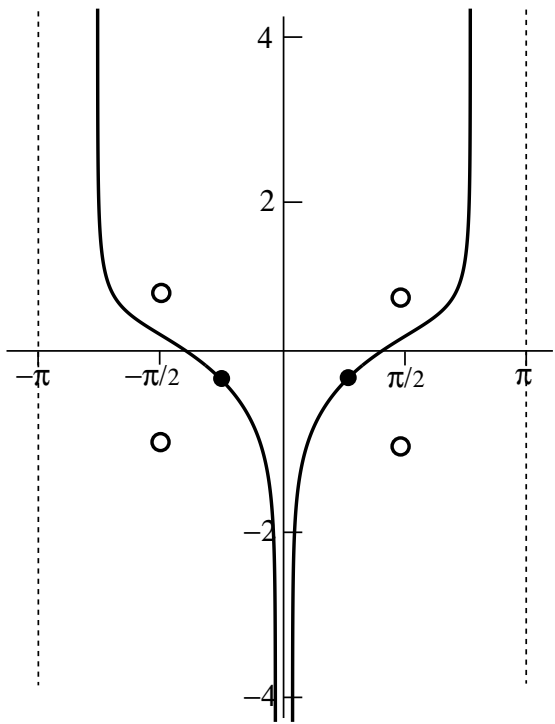}}
   \end{center}
  \end{minipage}
  \caption{{\small{The saddle-point contour for the integral in 
            \eqref{psiRint} for the oscillatory range 
            $0<\alpha < 1/\sqrt{2}.$  The interval $[-\pi,\pi]$ can be 
            replaced by a path that runs from $-\pi$ to $-\pi+i\infty$; 
            from that point through the saddle-point at the negative real 
            axis to $-i\infty$  and from that point through the saddle-point 
            at the positive real axis to $+\pi+i\infty$ and then to $+\pi.$ 
            Note that the contributions from the vertical half lines cancel 
            each other out.  On the contour shown, the imaginary part of the 
            phase function $-i(\omega_{\theta}-\theta \alpha)t$ is constant 
            (equal to $0,$ in fact). The real part tends to $-\infty$ in the 
            valleys at infinity, and has a maximum at the saddlepoints 
            (black dots).}}}\label{oscillint}
\end{figure}

The contour of integration goes through the saddle-point on the positive 
imaginary axis, that is, through 
$\theta_{\alpha}=  i\arccosh(\alpha/\sqrt{1-\alpha^{2}})$ and fixes
the imaginary part of $i \omega_{\theta}-i \theta \alpha.$  This is 
equivalent to fixing the real part of $ \omega_{\theta}-\theta \alpha.$  
We proceed as follows. Consider the integral along the 
contour in Figure~\ref{expodecint}.  We can make this into a 
closed contour by adding in segments from $-\pi \to \pi,$ 
$\pi \to \pi+i\infty$ and $-\pi+i\infty \to-\pi$, thus obtaining an 
integral over
\begin{equation}\label{closed}
  (-\pi,\pi) \cup (\pi, \pi + i \infty) \cup (\pi + i \infty, 
   \theta_{\alpha}) \cup (+ \theta_{\alpha},-\pi + i\infty) \cup
   (-\pi + i \infty, -\pi).
\end{equation}

\bigskip

The singular points of $\sqrt{1+\cos^2\theta} $ follow from solving 
$\cos^2 \theta=-1$, which gives $\theta=\pm\pi/2\pm i\arcsinh(1)$ (see the 
open dots in Figure~\ref{expodecint}).  We avoid the singularities and 
branch cuts of the square root in the integrand and of the function 
$\omega_\theta$ (these singularities are the same for both functions; 
see Figure~\ref{conformal}).  The  integrand is then analytic around and 
inside the contour in \eqref{closed}, so the integral around the contour 
is zero. The integrals over the curves indicated below are therefore equal,
\begin{equation}
  (-\pi,\pi) = - (\pi, \pi + i \infty) \cup (\pi + i \infty, 
   \theta_{\alpha}) \cup (+ \theta_{\alpha},-\pi + i\infty) \cup
   (-\pi + i \infty, -\pi).
\end{equation}

\bigskip

The steepest-descent curves for this integral are shown in 
Figure~\ref{expodecint}.  Furthermore the periodicity of the integrand 
modulo $2\pi$ means that the segment integrals from $\pi$ to 
$\pi + i \infty$ and from $-\pi + i \infty$ to $-\pi$ cancel, so these are 
not shown in Figure~\ref{expodecint}. 
Thus, the integral from $-\pi$ to $\pi$ equals the integral along the 
contour from $-\pi+i \infty$ to $\pi + i\infty$ through $\theta_{\alpha}.$
From Lemma~\ref{bound1} we conclude that
\begin{equation}\label{est}
  e^{-i(\omega_{\theta}-\alpha)t}={{\cal O}}\left(e^{-(1-\alpha) v t}\right)
\end{equation}
as $v\to+\infty$ in the strips $-\pi\le u <-\frac12\pi$ and 
$\frac12\pi<u\le\pi$. 
Hence, convergence at infinity on a contour as shown in 
Figure~\ref{expodecint} is guaranteed.

\subsection{Evaluating the main contribution}\label{main}

We evaluate $ e^{-i\omega_{\alpha}+i\theta_{\alpha}\alpha}$
for the saddle-point on the positive imaginary axis, that is, for 
\begin{equation}
  \theta_{\alpha}
  =i\arccosh\left(\frac{\alpha}{\sqrt{1-\alpha^{2}}}\right).
\end{equation}
Now, with $x=\alpha/\sqrt{1-\alpha^{2}}$, we obtain 
\begin{equation}\label{thesubalph}
  \theta_{\alpha}
  = i\arccosh \; x= 
  i\ln(x+(x^{2}-1)^{1/2})=i\ln\left(\frac{\alpha+\sqrt{2\alpha^{2}-1}}
  {\sqrt{1-\alpha^{2}}}\right).
\end{equation}
Thus 
\begin{equation}
  e^{i\theta_{\alpha}\alpha}
  =\exp\left(- \alpha\ln \left(\frac{\alpha +\sqrt{2 \alpha^{2}-1}}
   {\sqrt{1-\alpha^{2}}}\right)\right) = 
   \left(\frac{\sqrt{1-\alpha^{2}}}
   {\alpha +\sqrt{2\alpha^{2}-1}}\right)^{ \alpha}. 
\end{equation}

Let us now consider 
$\omega_{\alpha}
 =\arcsin\left(\frac{\sin \theta_{\alpha}}{\sqrt{2}}\right)$. 
We have
\begin{equation}
 \cos \theta_{\alpha}=\frac{\alpha}{\sqrt{1-\alpha^{2}}}, 
\end{equation}
so therefore 
\begin{equation} 
 \sin \theta_{\alpha}=\sqrt{1-\alpha^{2}/(1-\alpha^{2})}
  =\sqrt{\frac{1-2\alpha^{2}}{1-\alpha^{2}}}
  =i\sqrt{\frac{2\alpha^{2}-1}{1-\alpha^{2}}}.
\end{equation}
Thus we obtain 
\begin{equation} 
  \omega_{\alpha} =i\arcsinh\left(\frac{1}{\sqrt{2}}
    \sqrt{\frac{2\alpha^{2}-1}{1-\alpha^{2}}}\right).
\end{equation}
Now $\arcsinh\, x=\ln(x+\sqrt{x^{2}+1})$, so
\begin{align}
 \omega_{\alpha}&=i\ln \left(\frac{\sqrt{2\alpha^{2}-1}}
 {\sqrt{2}\sqrt{1-\alpha^{2}}}
 +\left(\frac{2\alpha^{2}-1}{2(1-\alpha^{2})}+1\right)^{1/2}\right)\nonumber \\
  &=i\ln\left(\frac{\sqrt{2\alpha^{2}-1}}{\sqrt{2}\sqrt{1-\alpha^{2}}}
+\frac{(2\alpha^{2}-1+2-2\alpha^{2})^{1/2}}
      {\sqrt{2(1-\alpha^{2})}}\right)\nonumber \\
  &=i\ln\left(\frac{1+\sqrt{2\alpha^{2}-1}}{\sqrt{2}\sqrt{1-\alpha^{2}}}
  \right).
\end{align}
Thus
\begin{equation}\label{expomega1}
 e^{-i\omega_\alpha}= \exp\left(\ln\frac{1+\sqrt{2\alpha^{2}-1}}
 {\sqrt{2}\sqrt{1-\alpha^{2}}}\right)=
 \frac{1+\sqrt{2\alpha^{2}-1}}{\sqrt{2}\sqrt{1-\alpha^{2}}}.
\end{equation}
and hence
\begin{equation}
 e^{-i\omega_{\alpha}+i\theta_{\alpha}\alpha}=
\left(\frac{\sqrt{1-\alpha^{2}}}{\alpha+\sqrt{2\alpha^{2}-1}}\right)^{\alpha}
\frac{1+\sqrt{2 \alpha^{2}-1}}{\sqrt{2}\sqrt{1-\alpha^{2}}}.
\end{equation}
Since $\omega_{\theta} -\theta_{\alpha} \alpha$ is an odd function of 
$\theta_{\alpha}$, we will obtain the reciprocal of this result for the 
saddle-point $-i\theta_\alpha.$ 
The saddle-point on the positive imaginary axis (when $\alpha> 1/\sqrt2$) 
is indeed the relevant one as we will see. 
The exact saddle-point contours are shown in Figures \ref{oscillint} and 
\ref{expodecint}. 

\bigskip

Now that we have established these preliminary results, we can proceed to 
prove
\begin{theorem}\label{newthm4}
If $1/\sqrt{2}+\varepsilon <|\alpha| <1-\varepsilon,$ then
\begin{equation}\label{Btildef}
 \psi_{R}(n,t)\sim 
 \frac{(-1)^{n+1} \left(\alpha+\sqrt{2\alpha^{2}-1}\right) t^{-1/2}}
      {\sqrt{2\pi (1-\alpha^{2}) \sqrt{2\alpha^{2}-1}}}
 \left(\left(\frac{\sqrt{1-\alpha^{2}}}{\alpha +
 \sqrt{2\alpha^{2}-1}}\right)^{\alpha}\frac{1+\sqrt{2\alpha^{2}-1}}
 {2\sqrt{1-\alpha^{2}}}\right)^{t}.
\end{equation}
\begin{equation}\label{psiLth4}
 \psi_{L}(n,t)\sim
 \frac{(-1)^{n}(1-\alpha) t^{-1/2}}
 {\sqrt{2\pi (1-\alpha^{2})\sqrt{2\alpha^{2}-1}}}
  \left(\left(\frac{\sqrt{1-\alpha^{2}}}{\alpha +
  \sqrt{2\alpha^{2}-1}}\right)^{\alpha}\frac{1+\sqrt{2\alpha^{2}-1}}
  {2\sqrt{1-\alpha^{2}}}\right)^{t}.
 \end{equation}
\end{theorem}

\begin{proof} 
We prove the equation for $\psi_{R},$ the proof of the equation for 
$\psi_{L}$ is very similar. Since 
$\omega_{\theta}=\arcsin\left(\frac{\sin \theta}{\sqrt{2}}\right),$ 
we have 
\begin{equation}
  \omega_{\theta}''
  =-\frac{\sin \theta}{\left(1+\cos^{2} \theta\right)^{3/2}}=
   -i(1-\alpha^{2})\sqrt{2\alpha^{2}-1}.    
\end{equation}
Since we know that $\cos^{2}\theta_{\alpha}=\alpha^{2}/(1-\alpha^{2}),$ 
we will obtain
\begin{equation}\label{cosnote}
  1+\cos^{2}\theta_{\alpha}=1/(1-\alpha^{2}),
\end{equation}
and we already have (see equation~\eqref{thesubalph}) 
\begin{equation}\label{expfrac}
  e^{i\theta_{\alpha}}
  =\frac{\sqrt{1-\alpha^{2}}}{\alpha+\sqrt{2\alpha^{2}-1}}.
\end{equation}
The standard formula from steepest descents tells us that 
\begin{equation}
  \psi_{R}(n,t) \sim \frac{(-1)^{n+1}}{2\pi}
  \frac{e^{-i\theta_{\alpha}-i(\omega_{\alpha}-\theta_{\alpha}\alpha)t}}
  {\sqrt{1+\cos^{2}\theta_{\alpha}}}
  \sqrt{\frac{2\pi}{t|\omega_{\alpha}^{''}|}}.
 \end{equation}
The theorem then follows by using equations \eqref{expomega1}, 
\eqref{psiLth4}, \eqref{cosnote} and \eqref{expfrac}.
\end{proof}

\bigskip
\noindent
\textbf{Remark}: 
The wave-mechanics calculation is conceptually much simpler than the 
path-integral analysis in Carteret {\emph{et al.}} \cite{3routes}. 
It is also simpler than the calculations of Chen and Ismail 
\cite{ChenIsmail}.

\section{Equivalence of the two approaches}\label{equivalence}

We have now completed the calculation begun in \cite{3routes} and obtained 
uniformly convergent asymptotics for the wavefunction via both methods. 
However, the functions $\psi_{L}$ and $\psi_{R}$ derived by each route did 
not appear to be the same.  If they really were different, this would be 
very alarming, as it would imply either that there is something wrong with 
Feynman's equivalence argument in \cite{FeynmanHibbs} or, more likely, that 
there was something wrong with our calculation!  

\bigskip

The first thing to note is that the quantity raised to the power $t$ in 
equation~\eqref{Btildef} dominates the asymptotics of the logarithm of the 
functions $\psi_{\rm{L}}$ and $\psi_{\rm{R}}$ from the Schr{\"o}dinger 
representation.  Let us call it $\tilde{B}(\alpha);$ that is, 
\begin{equation}\label{solution1}
  \tilde{B}(\alpha) =\left(\frac{\sqrt{1-\alpha^{2}}}
  {\alpha+\sqrt{2\alpha^{2}-1}}\right)^{\alpha}
  \frac{1+\sqrt{2\alpha^{2}-1}}{\sqrt{2}\sqrt{1-\alpha^{2}}}.
\end{equation}

These estimates agree with the asymptotics obtained using the method of 
Saff and Varga \cite{Saff} as used in \cite{3routes}, although this is 
not yet apparent.  According to the calculation in \cite{3routes}, the 
corresponding quantity from the path-integral representation, namely 
$B(\alpha),$ should be 
\begin{equation}\label{solution2}
2^{-\frac{\alpha}{2}} \times
\left(\frac{1+2\alpha-\sqrt{2\alpha^{2}-1}}{1+\alpha}\right)^{\alpha}
\left(\frac{\alpha^{2}+\sqrt{2\alpha^{2}-1}}{1-\alpha^{2}}
\right)^{(1-\alpha)/2},
\end{equation}
which would seem to be a different function. 

The demonstration of the equivalence to the result obtained by Saff and 
Varga's method needs some identities, starting with  
\begin{equation}\label{ident1}
 \left(1+\sqrt{2\alpha^{2}-1}\right)^{2}
 =2\left(\alpha^2 +\sqrt{2\alpha^2 -1}\right).
\end{equation}
Combining $1+2\alpha-\sqrt{2\alpha^2-1}$ with the other quantities is 
rather fiddly (see below).
To show that the two solutions \eqref{solution1} and \eqref{solution2} are 
equivalent, we will now employ the identity
\begin{equation}\label{ident2}
 \frac{1+2\alpha-\sqrt{2\alpha^2 -1}}{1+\alpha}
  = \frac{1+\sqrt{2 \alpha^2-1}}{\alpha + \sqrt{2\alpha^{2}-1}},
\end{equation}
which can easily be verified by cross-multiplication.
It follows immediately from this identity that
\begin{equation}
 \left(\frac{1+2\alpha-\sqrt{2\alpha-1}}{1+\alpha}\right)^{\alpha}=
 \left(\frac{1+\sqrt{2\alpha^{2}-1}}{\alpha+
   \sqrt{2\alpha^2-1}}\right)^{\alpha}
\end{equation}
and from equation \eqref{ident1} that
\begin{align}
   \left(\frac{1-\alpha^2}{\alpha^2+\sqrt{2\alpha^2-1}}
        \right)^{\frac{\alpha^2-1}{2}}
   &=\left(\frac{2(1-\alpha^2)}{(1+\sqrt{2\alpha^2-1})^2}
           \right)^{\frac{\alpha-1}{2}} \\
   &=\left(\frac{\sqrt{2}\sqrt{1-\alpha^2}}{1+\sqrt{2\alpha^2-1}}
           \right)^{\alpha}
    \frac{1+\sqrt{2\alpha^2-1}}{\sqrt{2}\sqrt{1-\alpha^{2}}}.
\end{align}

Thus
\begin{multline}
\left(\frac{1+2\alpha-\sqrt{2\alpha^2-1}}{1+\alpha}\right)^{\alpha}
 \left(\frac{1-\alpha^2}{\alpha^2-\sqrt{2\alpha^2-1}}
       \right)^{\frac{\alpha-1}{2}}   \\
 = 2^{\alpha/2}\left(\frac{\sqrt{1-\alpha^2}}{\alpha + \sqrt{2\alpha^2-1}}
  \right)^{\alpha}\frac{1+\sqrt{2\alpha^2-1}}{\sqrt{2}\sqrt{1-\alpha^2}}
\end{multline}
so that the formul\ae\, for $\tilde{B}(\alpha)$ and $B(\alpha),$ in 
\eqref{solution1} and \eqref{solution2} respectively are indeed equivalent.

\bigskip

In fact, the representations for the two functions are completely 
equivalent, but some of the steps required to prove this need some rather 
subtle calculations involving special functions, as we will now demonstrate.  
These technical lemmas will also allow us to rederive the symmetry relations 
in the wave-mechanics picture that were first proved (for the path-integral 
picture) in \cite{Ambainis01}. 
For convenience we recall the integral representations of Theorem~\ref{NV1}
\begin{equation}\label{intpsiL}
 \psi_{L}(n,t)= \frac{1}{2\pi} 
 \int_{-\pi}^{\pi}\left(1+\frac{\cos 
 \theta}{\sqrt{1+\cos^{2}\theta}}\right)
 e^{-i(\omega_{\theta}+\theta \alpha )t}\; d\theta, 
\end{equation}
\begin{equation}\label{intpsiR}
  \psi_{R}(n,t)=\frac{1}{2\pi} 
  \int_{-\pi}^{\pi} \frac{e^{i \theta}}{\sqrt{1+\cos^{2} 
  \theta}}e^{-i(\omega_{ \theta}+\theta \alpha )t}\; d\theta.
\end{equation}
where $\alpha=n/t$.  Note that we have omitted the factors 
$\frac{1+(-1)^{n+t}}{2}$ since $n$ and $t$ must have the same parity because 
the walker must move at each time-step.

From these integral representations obtained in the Schr{\"{o}}dinger 
picture \cite{NayaknV}, we prove in this section the symmetry relations for 
$\psi_{L}$ and $\psi_{R}$ and the relations first proved for the Jacobi 
polynomials. 
These results could previously only be proved in the path-integral picture 
\cite{Ambainis01}.

\subsection{Symmetry properties}\label{symmetries}

\begin{lemma}\label{lempsir}
The function $\psi_{R}$ of \eqref{intpsiR} satisfies the symmetry relation
\begin{equation}
  \label{psiRsym}
\psi_{R}(-n,t) =
(-1)^{n+1} \psi_{R}(n+2,t).
\end{equation}
\end{lemma}
This is the symmetry relation in Theorem~\ref{Amb2} for $\psi_{R}(n,t).$
\begin{proof}
We have, from equation \eqref{intpsiR}
\begin{equation}\label{psir}
\psi_{R}(n,t) =
\frac1{\pi}\int_{0}^\pi \frac{\cos((n-1)\theta)
 \cos(\omega_\theta t)}
 {\sqrt{1+\cos^2\theta}}\,d\theta-\frac1{\pi}\int_{0}^\pi 
 \frac{\sin((n-1)\theta)\sin(\omega_\theta t)}
 {\sqrt{1+\cos^2\theta}}\,d\theta.
\end{equation}
The first integral vanishes when $n$ is even, the second one when $n$ is odd.
To verify this, split $[0,\pi]=[0,\frac12\pi]\cup[\frac12\pi,\pi]$ and 
write on the second interval $\theta=\pi-\theta'$.
We conclude that $\psi_{R}(1-n,t)=\psi_{R}(1+n,t)$ when $n$ is odd, and
$\psi_{R}(1-n,t)=-\psi_{R}(1+n,t)$ when $n$ is even. This proves the lemma.
\end{proof}

\begin{lemma}\label{lempsil}
The function $\psi_{L}$ of equation \eqref{intpsiL} satisfies the symmetry 
relation
\begin{equation}\label{psilsym}
(t-n)\psi_{L}(-n,t)=(-1)^n(t+n)\psi_{L}(n,t).
 \end{equation}
\end{lemma}

\noindent
This is the symmetry relation in Theorem~\ref{Amb2} for $\psi_{L}(n,t).$
\begin{proof}
We have from \eqref{intpsiL}
\begin{equation}\label{psil}
\psi_{L}(n,t) =
\widetilde{\psi}_{L}(n,t)+\frac1{2\pi}\int_{-\pi}^\pi \frac{\cos\theta}
 {\sqrt{1+\cos^2\theta}} e^{-i(t\omega_\theta +n\theta)}\,d\theta,
\end{equation}
where
\begin{align}
 \widetilde{\psi}_{L}(n,t) &=\frac{1}{2\pi}\int_{-\pi}^{\pi}
   e^{-i(t\omega_{\theta}+n\theta)}\, d\theta   \label{psitilde1}\\
   &=\frac{1}{\pi}
 \int_{0}^{\pi}\cos(n\theta)\cos(t\omega_{\theta})\; 
   d\theta -\frac{1}{\pi}\int_{0}^{\pi}\sin(n\theta)
   \sin(t\omega_{\theta}) \, d\theta. \label{psitilde2}
\end{align}
The first integral vanishes when $n$ is odd, the second one when $n$ is even.
This gives the symmetry relation
\begin{equation}\label{psiLtsym}
 \widetilde{\psi}_{L}(-n,t)=(-1)^{n} \widetilde{\psi}_{L}(n,t).
 \end{equation}
Next observe that 
\begin{equation}\label{omder}
\frac{d\omega_\theta}{d\theta}=\frac{\cos\theta}
 {\sqrt{1+\cos^2\theta}},
 \end{equation}
and that an integration by parts in the integral in equation \eqref{psil} 
gives us that
\begin{equation}\label{psilr}
\psi_{L}(n,t)=\frac{t-n}{t}\widetilde{\psi}_{L}(n,t).
 \end{equation}
Finally we use equations \eqref{psiLtsym} and \eqref{psilr} to  
complete the proof of the lemma.
\end{proof}

\smallskip

\noindent
{\bf{ Remark:}}
The symmetry relations for $\psi_{L}$ and $\psi_{R}$ also follow from the
following property of the Jacobi polynomials:
\smallskip
\begin{equation}\label{jacsym}
   \binom{m}{\ell}J_{m}^{(u,-\ell)}(x)=
   \binom{m+u}{\ell}\left(\frac{1+x}{2}\right)^{\ell}
   J_{m-\ell}^{(u,\ell)}(x) ,\quad 0 \le \ell \le m.
\end{equation}
This formula follows from the representation of the Jacobi polynomial in 
terms of the hypergeometric function (cf. \cite[p. 151, (6.35)]{temme96}) 
in combination with a functional relation of this function (third line of 
equation (5.5) in \cite[p. 110]{temme96}).  
The result in \eqref{jacsym} combines the first case in \eqref{psiRJac} 
with the second case, and it also implies the symmetry rule for $\psi_{R},$ 
and similarly for \eqref{psiLJac}.

\subsection{The ${\boldsymbol{\psi}-}$functions in terms of 
Jacobi polynomials}\label{psijac}

We will now prove that the $\psi-$functions with the integral 
representations given in \eqref{intpsiL} and \eqref{intpsiR} can be written 
in terms of the Jacobi polynomials as  in Theorem~\ref{Amb2}.
This step will require the use of generating functions.  These provide a 
method for writing a series as the coefficients of a formal power series, 
in a dummy variable, $z,$ for ease of manipulation. The powers of $z$ are 
then the summation variable for the series.   For a basic introduction to 
the theory of generating functions, see \cite{Wilf}; for an advanced 
treatment, see \cite{Jackson}.

We will use these generating functions to give us an {\emph{exact}} 
representation of the $\Psi$ functions as functions of $t,$ thus each 
labelled term in the series will be the function for that value of $t.$  
Therefore, our approach will be based on generating functions that contain 
the $\psi-$functions with $t$ as the summation variable. 
We only consider sums of $\psi-$functions with $n$ and $t$ having the 
same parity.

\subsection{Some generating functions for 
${\boldsymbol{\psi}}$}\label{evaluating}

For convenience, we will define the following generating functions, 
which can be obtained from the Schr{\"o}dinger representation of 
the wavefunction.

\begin{theorem}\label{genthm}
Consider the generating functions for $|z|<1$:
\begin{align}%\label{genfunser}
F_m(z)&=\sum_{t=0}^\infty \psi_{R}(2m+1,2t+1) z^t, \label{fmser}      \\
G_m(z)&=\sum_{t=0}^\infty \psi_{R}(2m,2t) z^t,     \label{gmser}       \\
H_m(z)&=\sum_{t=0}^\infty \widetilde\psi_{L}(2m+1,2t+1) z^t, \label{hmser} \\
I_m(z)&=\sum_{t=0}^\infty \widetilde\psi_{L}(2m,2t) z^t,\label{imser} 
\end{align}
where $\widetilde\psi_{L}(n,t)$ is defined in \eqref{psitilde1}.
After the summations have been performed these functions become, 
respectively,  
\begin{align}
  F_m(z)&=\frac{2^{m-\frac12} z^m}{\sqrt{1+z^2}(1-z+\sqrt{1+z^2})^{2m}}, 
        & m=0,1,2,\ldots,  \label{fmres}    \\
  G_m(z)&=-\frac{2^{m-1} z^m}{\sqrt{1+z^2}(1-z+\sqrt{1+z^2})^{2m-1}},  
        & m= 1,2,3\ldots,     \label{gmres}    \\
  G_0(z)&=\frac{ z}{\sqrt{1+z^2}(1-z+\sqrt{1+z^2})},    
        &   \label{g0res}    
\end{align}
\begin{align}
  H_m(z)&=\frac{2^{m-\frac12}(1+z)z^m}{\sqrt{1+z^2}(1-z+\sqrt{1+z^2})^{2m+1}},
        & m=0,1,2,\ldots\label{hmres} \\
  I_m(z)&=\frac{2^{m-1}(1+z) z^m}{\sqrt{1+z^2}(1-z+\sqrt{1+z^2})^{2m}},
        & m=1,2,3\ldots,\label{imres} \\
  I_0(z)&=\frac{1}{\sqrt{1+z^2}}-\frac{z}{\sqrt{1+z^2}(1-z+\sqrt{1+z^2})}.
        & \label{i0res} 
\end{align}
\end{theorem}
These summations can be done using some fiddly, but essentially mechanical 
manipulations; we have included detailed proofs for $F_m(z)$ and a sketch 
of that for $G_m(z)$ in an appendix, in Subsection~\ref{construction}.  
The other generating functions can be obtained via similar constructions. 

\bigskip

\noindent
For $H_m(z)$ we obtain
\begin{align}\label{hmzeq}
 H_m(z) 
  &=
\frac1{2\pi}\int_{-\pi}^\pi {e^{-i((2m+1)\theta+\omega_\theta)}}
  \,\frac{1}
 {1-ze^{-2i\omega_\theta}}\,d\theta\\
  &=\frac{1+z}{2\pi\sqrt{2}}\int_{0}^\pi 
\frac{\cos ((2m+2)\theta)-\cos(2m\theta)}
 {1-2z\cos2\omega_\theta+z^2}\,d\theta\\
  &=\frac{1+z}{2(1-z)}\left[F_{m+1}(z)-F_m(z)\right]. 
\end{align}
Taken together with \eqref{fmres}, this becomes equation \eqref{hmres}. 

\bigskip

\noindent
For $I_m(z)$ we have a useful intermediate result, namely that 
\begin{equation}\label{imzeq}
I_m(z) =\frac{\sqrt{2}}{4(1-z)}
        \left[2(2-z)F_{m}(z)-zF_{|m-1|}(z)-zF_{m+1}(z)\right]. 
\end{equation}
%\end{proof}

It remains to compare these with some generating functions for Jacobi 
polynomials. 
For an introduction to the theory of Jacobi polynomials we refer the 
reader to \cite{Olver,AskeyRoy} and~\cite{temme96}.  
This correspondence between the two sets of generating functions forms 
the central plank of the Feynman equivalence between the two 
representations of the wave-function for this system.

\subsection{Comparing the generating functions for 
$\boldsymbol{\psi}$}\label{comparing}

We now compare the generating functions \eqref{fmres} -- \eqref{i0res} 
from the wave-mechanics representation, with the generating function of 
the Jacobi polynomials (cf. \cite[p. 298]{AskeyRoy}) from the path-integral 
representation \cite{3routes} to complete our proof of the equivalence of 
the two approaches:
\begin{equation}\label{jpsum}
  \sum_{k=0}^\infty J_k^{(r,s)}(x)z^k=
  \frac{2^{r+s}}{R(1-z+R)^{r}(1+z+R)^{s}}, \quad |z|<1,
\end{equation}
where $R=\sqrt{1-2xz+z^2}$, which for $x=0$ becomes 
\begin{equation}\label{jpsum0}
  \sum_{k=0}^\infty J_k^{(r,s)}(0)z^k=
  \frac{2^{r+s}}{\sqrt{1+z^2}(1-z+\sqrt{1+z^2})^{r}
  (1+z+\sqrt{1+z^2})^{s}}.
\end{equation}
By applying the Cauchy integral formula we find that 
\begin{equation}\label{jaccauch}
  J_k^{(r,s)}(0)=\frac{2^{r+s}}{2\pi i}\oint 
  \frac{dz}{\sqrt{1+z^2}(1-z+\sqrt{1+z^2})^{r}
  (1+z+\sqrt{1+z^2})^{s}z^{k+1}},
\end{equation}
where the integral is taken over a circle with radius less than unity.

\begin{lemma}\label{lemmapsiR}
For $n=0$ we have
\begin{align}\label{psir0tjac}
  \psi_R(0,0) &= 0, \\ 
  \psi_R(0,2t) &= \frac12 J_{t-1}^{(1,0)}(0)=\frac12 
  (-1)^{t-1}J_{t-1}^{(0,1)}(0),
  \quad t=2,4,6,\ldots\, .
 \end{align}
For  $0< n \le t$, $n$ and $t$ having the same parity, we have
\begin{equation}\label{psirjac}
\psi_{R}(n,t) =
 2^{-\frac{n}{2}}(-1)^{\frac{t-n}{2}}(-1)^{n+1}
   J_{\frac{t-n}{2}}^{(0,n-1)}(0).
\end{equation} 
\end{lemma}
\noindent
This gives the first case of equation \eqref{psiRJac} from the 
wave-mechanics representation, which was originally obtained via the 
path-integral method.

\begin{proof}
From \eqref{fmser} and \eqref{fmres} we find, as in \eqref{jaccauch},
\begin{equation}\label{fmcauch}
  \psi_R(2m+1,2t+1)=\frac{2^{m-\frac12}}{2\pi i}\oint 
  \frac{ z^m}{\sqrt{1+z^2}(1-z+\sqrt{1+z^2})^{2m}}\,\frac{dz}{z^{t+1}}.
\end{equation}
We now compare \eqref{jaccauch} with \eqref{fmcauch} and take $u=2m,$ 
$v=0,$ and $s=t-m.$  This gives
\begin{equation}\label{psiroddjac1}
  \psi_R(2m+1,2t+1)=2^{-m-\frac12} J_{t-m}^{(2m,0)}(0),\quad 0\le m\le t.
 \end{equation}
Using the symmetry rule for the Jacobi polynomials
\begin{equation}\label{jsym}
 J_n^{(r,s)}(-x)=(-1)^nJ_n^{(s,r)}(x),
 \end{equation}
which follows from \eqref{jpsum} by putting $x\to-x, z\to-z$, we find
\begin{equation}\label{psiroddjac2}
  \psi_R(2m+1,2t+1)=2^{-m-\frac12} (-1)^{t-m} J_{t-m}^{(0,2m)}(0),
       \quad 0\le m\le t.
 \end{equation}
For the even case, we obtain from \eqref{gmser}, \eqref{gmres}, and 
\eqref{jsym}
\begin{equation}\label{psirevenjac}
  \psi_R(2m,2t)=-2^{-m} J_{t-m}^{(2m-1,0)}(0)
               =-2^{-m} (-1)^{t-m} J_{t-m}^{(0,2m-1)}(0),\quad 0 < m \le t,
\end{equation}
and from \eqref{gmser} and \eqref{g0res} we obtain \eqref{psir0tjac}.
Combining \eqref{psiroddjac2} and \eqref{psirevenjac} in one formula 
gives \eqref{psirjac}.  This proves the lemma.
\end{proof}

\begin{lemma}\label{lemmapsiL}
For $n=0$  we have
\begin{align}\label{psil0tjac}
 \psi_{L}(0,0) &= 1, \\
 \psi_{L}(0,t) &= J_{t/2}^{(0,0)}(0)-\frac12 J_{t/2-1}^{(1,0)}(0), 
\quad t=2,4,6,\ldots\, .
\end{align}
For $0< n < t$, $n$ and $t$ having the same parity, we have
\begin{equation}\label{psiljac}
\psi_{L}(n,t)
=2^{-n/2-1}(-1)^{(t-n)/2-1}J_{(t-n)/2-1}^{(1,n)}(0).
\end{equation}
\end{lemma}
\noindent
This gives the first case of \eqref{psiLJac} from the path-integral 
representation, now obtained via wave-mechanics.

\bigskip

\begin{proof}
We obtain from \eqref{hmser} and \eqref{hmres}
\begin{align}
\widetilde{\psi}_{L}(2m+1,2t+1)
&=2^{-m-\frac32}
 \left[J_{t-m}^{(2m+1,0)}(0)+
   J_{t-m-1}^{(2m+1,0)}(0)\right]\label{wtpsirjac1}\\
   &=2^{-m-\frac32}\frac{2t+1}{t+m+1}
   J_{t-m}^{(2m+1,-1)}(0),\label{wtpsirjac2}
\end{align}
where we have used the relation for the Jacobi polynomials 
(cf. \cite[p.782, (22.7.19)]{AnStegun})
\begin{equation}\label{jaceq}
  (u+v+2k)J_k^{(u,v-1)}(x)
  =(u+v+k)J_k^{(u,v)}(x)
  +(u+k)J_{k-1}^{(u,v)}(x).
\end{equation}
We use \eqref{jacsym}, \eqref{psilr} and \eqref{jsym}, and obtain
\begin{equation}\label{psiloddjac}
\psi_{L}(2m+1,2t+1)
    =2^{-m-\frac32}(-1)^{t-m-1}J_{t-m-1}^{(1,2m+1)}(0).
\end{equation}
For the even case we obtain from \eqref{imser} and \eqref{imres}
\begin{align}
\widetilde{\psi}_{L}(2m,2t)
        &=2^{-m-1}\left[J_{t-m}^{(2m,0)}(0)
          +J_{t-m-1}^{(2m,0)}(0)\right]\label{wtpsileven1} \\
        &=2^{-m-1}(-1)^{t-m-1}\frac{t}{t-m}
          J_{t-m-1}^{(1,2m)}(0),\label{wtpsileven2}
\end{align}
where we used \eqref{jaceq}, \eqref{jacsym} and \eqref{jsym}.  By using 
\eqref{psilr} we obtain
\begin{equation}\label{psilevenjac}
   \psi_{L}(2m,2t)
   =2^{-m-1}(-1)^{t-m-1}J_{t-m-1}^{(1,2m)}(0),\quad m=1,2,3,\ldots\,.
\end{equation}
From \eqref{imser} and \eqref{i0res} we obtain \eqref{psil0tjac}.
Combining \eqref{psiloddjac} and \eqref{psilevenjac} into a single formula  
gives equation \eqref{psiljac}. This proves the lemma.
\end{proof}

\noindent
As we have now established that both sets of generating functions match, 
this completes the proof of the equivalence of the results obtained via 
the path-integral and wave-mechanics representations.  It should be noted 
that we have proved the two representations are exactly equivalent 
{\emph{for all time,}} as opposed to being only asymptotically equivalent 
in the long-time limit. It is one of the curious features of generating 
function methods that they can be used to prove the existence of a 
one-to-one correspondence between the two sets (i.e., our $\psi$-functions) 
counted by the two series, without actually finding the explicit bijection.

\subsection{Summary of the equivalence results}\label{summary}

While the coined quantum walk can be thought of as a quantum analogue 
of the discrete-time classical random walk \cite{Watrouswalk,Aharonov00}, 
it should be noted that the quantum model inherits its discrete time 
parameter {\emph{directly}} from the classical model; the discreteness 
was not introduced by hand as part of the quantization procedure.  Also, 
we have not defined a Hamiltonian for this system at all, so the problem 
of ambiguities in the time derivatives of the action does not arise.  We 
have now shown the full Feynman equivalence for this system, though some 
results seem easier to derive in one approach than in the other.

\begin{itemize}
\item 
We have obtained the symmetry rules directly from the integral 
representations for the $\psi$-functions. It is not necessary to represent 
the $\psi$-functions as Jacobi poynomials and then use the symmetry 
properties of Jacobi polynomials (as was done in \cite{Ambainis01}) to 
obtain this result.

\item 
The relations between the $\psi$-functions and the Jacobi polynomials have 
been obtained directly from the integral representations, though we needed 
to develop some technical tools for this.  These consisted of a few extra 
properties of the Jacobi polynomials, and the generating functions containing 
the $\psi$-functions. 
The proofs using these generating functions are conceptually straightforward, 
although the details are quite technical.  The Feynman path-integral approach 
of \cite{Ambainis01} (which is a finite sum here) would seem to be simpler 
if these relations are all one wants.

\item 
We were able to establish some new expressions for the values of the two 
components of the $\psi$-function at $n=0$ as a function of time, in 
equations \eqref{psir0tjac} and \eqref{psil0tjac}, which were not known 
before.

\end{itemize}

\section{Physical interpretation of these results}\label{interpretation}

Now that we have established that the rather counter-intuitive results 
obtained from the wave-mechanics picture really are equivalent to the 
Airy functions obtained from the path-integral approach, we are left with 
the little mystery of their physical interpretation.  In this region of 
exponential decay these waves have complex wavenumbers. This phenomenon is  
called evanescence.   And herein lies the mystery; the conventional wisdom 
is that evanescent waves are only ever seen in the presence of absorbing 
media, such as light waves being absorbed into a conducting surface, but 
there is no such surface here and the evolution is {\emph{unitary}}, by 
the initial assumptions that went into constructing the model. 
In fact, the phenomenon of evanescence is rather more widespread; it occurs 
in a great many systems if you know where to look.   In a pioneering paper 
in the early 1990s, Michael Berry showed that evanescent behaviour is much 
more common than had been previously thought, after being inspired by some 
work by Aharonov, Anandan, Popescu and Vaidman in \cite{timetrans}.  Berry 
gave a detailed discussion of how this phenomenon occurs in optics, at the 
edges of the almost ubiquitous ``Gaussian'' beams in \cite{evanescent}. 

The wavefunction for this system tends to an Airy function in the asymptotic 
limit, as was proved analytically in \cite{3routes}.  We evaluated the 
integrals in the path-integral picture using the method of steepest 
descents, which in this case featured a pair of coalescing saddle-points 
\cite{3routes,Wongpaper}.  
Since then, various authors have discussed the connection between the 
discrete walk on the infinite line and interference phenomena in the quantum 
optics of dispersive media.  This connection was first described by Knight, 
Roldan and Sipe in a series of papers \cite{origininf,opticimpl,interfwalk}
and further clarified by Kendon and Sanders in \cite{complementarity}. 

As we enter the exponential decay region, the two original stationary points 
of the phase function merge and then two new saddle-points are born, which 
move off the real axis as a complex conjugate pair.  The behaviour of the 
momentum closely follows that of $\omega_{\theta},$ which was plotted in 
Figure~\ref{omegatrack}.  
In the wave-mechanics picture, we found that the momentum becomes purely 
imaginary in the exponential decay region; indeed, the techniques we used to 
evaluate the integral relied on this fact.  So, the behaviour of the walk in 
the exponential decay region is a pure exponential decay; there is no 
oscillatory behaviour.   

Within the interpretation begun by Knight {\emph{et al.,}} it was first 
suggested to us by Achim Kempf \cite{Achim} that the specific evanescence 
phenomena that we have discussed in this paper are analogues of what are 
called the Sommerfeld and Brillouin precursors.  
Specifically, the exponential decay region can be identified with the 
{\bf{Sommerfeld precursor}} 
(see for example, \cite{Sommerfeld}) 
and the distinctive peaks in the probability distribution would be an 
example of the {\bf{Brillouin Precursor}}
(see for example \cite{Brillouin}). 
Our results here and the previous results in \cite{3routes} provide the 
first analytic evidence for this identification.

\section{Conclusions}\label{conclusion}

In this paper we have completed the analysis begun in \cite{3routes}, 
thus meeting the challenge made in \cite{Ambainis01} to prove all their 
theorems about the unrestricted quantum walk on the line in both the 
path-integral and wave-mechanics representations.  We have also proved 
some additional identities that we believe to be novel.  In the course 
of doing this, we have had to generalise the method of stationary phase 
in a way that may have applications beyond this problem.  We have also 
proved the exact Feynman equivalence between the two representations
directly, by reducing the problem to purely combinatorical constructions 
which may also be of wider interest.

Lastly, we have supplied a physical interpretation for our results, in terms 
of certain evanescent phenomena from the quantum optics of dispersive media. 
This interpretation is a somewhat counter-intuitive one, as it would seem to 
require an effective dissipation that acts on the walker in a way that is 
analogous to the effect of a dielectric medium on light, despite the fact 
that the evolution of the system is unitary by assumption.

%---------------------%
\subsection*{Acknowledgments}
%---------------------%

HAC would like to acknowledge some inspiring conversations with 
Sir Michael Berry, Mourad Ismail and Achim Kempf.  HAC was supported 
by MITACS, and would like to thank the Perimeter Institute and 
the IQC at the University of Waterloo for hospitality. 
LBR would also like to thank Ashwin Nayak for some interesting 
conversations.  
The research of LBR was partially supported by an NSERC operating grant.  
NMT acknowledges financial support from Ministerio de Educaci\'on y 
Ciencia (Programa de Sab\'aticos) from project SAB2003-0113.

%---------------------%
%\end{acknowledgments}
%---------------------%
%%%%%%%%%%%%%%%%%%%%%%%%%%%%%%%%%%%%%%%%%%%%%%%%%%%%%%
%-------------------------------------%
%%%% construct refs with natbib... %%%%
%\bibliographystyle{amsplain} %%%
%\bibliography{qwalksg} %Refs numbered in the order they appear in text
%----------------%

%-------------------------------------%
%%%% construct refs with BibTeX... %%%%
\bibliographystyle{amsplain} %Refs numbered in alphabetical order
%%%\bibliographystyle{alpha} %Refs in alphabetical order, alphanumeric tag.
%\bibliography{qwalksg}
%-------------------------------------%

%-------------------------------------------%
%%% ...then replace with .bbl file when finished:

\providecommand{\bysame}{\leavevmode\hbox to3em{\hrulefill}\thinspace}
\providecommand{\MR}{\relax\ifhmode\unskip\space\fi MR }
\providecommand{\MRhref}[2]{%
  \href{http://www.ams.org/mathscinet-getitem?mr=#1}{#2}
}
\providecommand{\href}[2]{#2}

%%%%---------------------------%

%%%Appendix begins here%%%%%%%%%%%%%%%%%%

\section{Appendices}

\subsection{Construction of the generating functions}\label{construction}

Here we give the details of the construction for the generating functions 
from Subsection~\ref{evaluating}.

\subsubsection{Proof of the construction for $\boldsymbol{F_m(z)}$}

\begin{proof}
We give a detailed proof for $F_m(z)$.
We substitute equation \eqref{intpsiR} into equation \eqref{fmser} and obtain 
\begin{multline}\label{fnzeqA}
 F_m(z) = 
 \frac1{2\pi}\int_{-\pi}^\pi \frac{e^{-i(2m\theta+\omega_{\theta})}}
 {\sqrt{1+\cos^2\theta}}\,\frac{1}
 {1-ze^{-2i\omega_{\theta}}}\;d\theta\\
 =\frac1{2\pi}\int_{-\pi}^\pi \frac{e^{-i(2m\theta+\omega_{\theta})}}
 {\sqrt{1+\cos^2\theta}}\,
   \frac{1-z\cos2\omega_{\theta}-iz\sin2\omega_{\theta}}
 {1-2z\cos2\omega_{\theta}+z^2}\;d\theta\\
 =\frac1{2\pi}\int_{-\pi}^\pi 
 \frac{\{\cos(2m\theta+\omega_{\theta})-i\sin(2m\theta+\omega_{\theta})\}
 \{1-z\cos2\omega_{\theta}-iz\sin2\omega_{\theta}\}}
 {\sqrt{1+\cos^2\theta}\,(1-2z\cos2\omega_{\theta}+z^2)}\;d\theta.
\end{multline}
Since $\omega_{\theta}$ is an odd function, the imaginary parts will 
vanish. The interval $[-\pi,\pi]$ can be reduced to $[0,\pi]$ and we obtain
\begin{align}\label{fnzeqB1}
 F_m(z)&=\frac1{\pi}\int_{0}^\pi \frac{\cos(2m\theta+\omega_{\theta})
   (1-z\cos2\omega_{\theta})-z\sin2\omega_{\theta}
   \sin(2m\theta+\omega_{\theta})}{\sqrt{1+\cos^2\theta}\,
   (1-2z\cos2\omega_{\theta}+z^2)}\;d\theta\\
  &= \frac1{\pi}\int_{0}^\pi \frac{\cos(2m\theta+\omega_{\theta})}
  {\sqrt{1+\cos^2\theta}\,(1-2z\cos2\omega_{\theta}+z^2)}
  \;d\theta \nonumber \\
 &\quad \quad -\frac{z}{\pi}\int_{0}^\pi \frac{\cos(2m\theta+\omega_{\theta})
  \cos (2\omega_{\theta})+\sin(2m\theta+\omega_{\theta})\sin2\omega_{\theta}}
  {\sqrt{1+\cos^2(\theta)}\,(1-2z\cos2\omega_{\theta}+z^2)}\;d\theta.
\end{align}
The numerator of the second integral can be written as 
$\cos(2m\theta-\omega_{\theta})$, and it follows that
\begin{align}\label{fnzeqB2}
 F_m(z)&=\frac1{\pi}\int_{0}^\pi \frac{\cos(2m\theta+\omega_{\theta})}
 {\sqrt{1+\cos^2\theta}\,(1-2z\cos2\omega_{\theta}+z^2)}\;d\theta \nonumber \\
 &\quad \quad -\frac{z}{\pi}\int_{0}^\pi \frac{\cos(2m\theta-\omega_{\theta})}
  {\sqrt{1+\cos^2\theta}\,(1-2z\cos2\omega_{\theta}+z^2)}\;d\theta. 
\end{align}
We can now use simple trigonometric identities for the cosines in the 
numerators, to obtain
\begin{align}\label{fnzeqB3}
 F_m(z) &=\frac1{\pi}\int_{0}^\pi \frac{\cos 2m\theta\cos\omega_{\theta}-
  \sin2m\theta\sin\omega_{\theta}}
  {\sqrt{1+\cos^2\theta}\,(1-2z\cos2\omega_{\theta}+z^2)}\,
 \,d\theta \nonumber \\
 &\quad \quad -\frac{z}{\pi}\int_{0}^\pi \frac{
 \cos 2m\theta\cos\omega_{\theta}+\sin2m\theta\sin\omega_{\theta}} 
 {\sqrt{1+\cos^2\theta}\,(1-2z\cos2\omega_{\theta}+z^2)}\;d\theta.
\end{align}
Now we use $\sin\omega_{\theta}=(\sin \theta)/\sqrt2$, and observe 
that the terms with the sine functions do not contribute to the integrals; 
this follows easily by performing the transformation 
$\theta=\theta^\prime+\frac12\pi$.  Using also 
$\cos\omega_\theta=\sqrt{\frac12(1+\cos^2\theta)}$, we obtain
\begin{equation}
  \label{fnzeq2}
 F_m(z)=\frac{1-z}{\pi\sqrt{2}}\int_{0}^\pi \frac{\cos (2m\theta)}
 {1-2z\cos2\omega_\theta+z^2}\,d\theta. 
\end{equation}
For the final step we use formula (3.615) (1) of Gradshteyn and Ryzhik 
\cite{grad}, that is,
\begin{equation}
\int_{0}^{\frac12\pi} \frac{\cos (2m\theta)}
 {1-a^2\sin^2\theta}\,d\theta=\frac{(-1)^m\pi}{2\sqrt{1-a^2}}
 \frac{(1-\sqrt{1-a^2})^{2m}}{a^{2m}}, \quad |a^2|<1, 
      \quad m=0,1,2,\ldots\;.
\end{equation}
We observe that $\cos2\omega_\theta=\cos^2\theta$, and take $a^2=-2z/(1-z)^2$. 
This gives the expression in \eqref{fmres}, as advertised.

\subsubsection{Outline of the proof for $\boldsymbol{G_m(z)}$}

The proof of equation \eqref{gmres} for $G_m(z)$ uses essentially the 
same manipulations:
\begin{align}\label{gmzeq}
G_m(z) &=
\frac1{2\pi}\int_{-\pi}^\pi \frac{e^{i(1-2m)\theta}}
{\sqrt{1+\cos^2\theta}}\,\frac{1}
{1-ze^{-2i\omega_{\theta}}}\;d\theta\\
 &=\frac1{2\pi}\int_{-\pi}^\pi
 \frac{\cos(2m-1)\theta(1-z\cos2\omega_{\theta})-
 z\sin(2m-1)\theta\,\sin2\omega_{\theta}}
 {\sqrt{1+\cos^2\theta}\,(1-2z\cos2\omega_{\theta}+z^2)}\;d\theta.
\end{align}
The second integral can be broken into two terms as follows:
\begin{align}
  G_m(z) &=\frac1{\pi}\int_{0}^\pi \frac{\cos(2m-1)\theta}
    {\sqrt{1+\cos^2\theta}\,
    (1-2z\cos2\omega_{\theta}+z^2)}\;d\theta\\
  &\quad
    -\frac{z}{\pi}\int_{0}^\pi
    \frac{\cos(2m-1)\theta\cos2\omega_{\theta}+
    \sin(2m-1)\theta\sin2\omega_{\theta}}{\sqrt{1+\cos^2\theta}\,
    (1-2z\cos2\omega_{\theta}+z^2)}\;d\theta.
\end{align}
The first integral vanishes (substitute $\theta=\theta^\prime+\frac12\pi$). 
The same holds for the contributions from the cosine terms in the second 
integral. This gives
\begin{equation}
   G_m(z) = -\frac{z}{\pi}\int_{0}^\pi
   \frac{\sin(2m-1)\theta\sin\theta}{
   1-2z\cos2\omega_{\theta}+z^2}\;d\theta.
\end{equation}
Using the fact that
\begin{align}
 \cos 2m\theta &= \cos(2m-1)\theta \cos\theta+\sin(2m-1)\theta \sin\theta \\
 \cos (2m-2)\theta &= \cos(2m-1)\theta \cos\theta-\sin(2m-1)\theta \sin\theta
\end{align}
we can then write
\begin{equation}
 G_m(z) = \frac{z}{2\pi}\int_{0}^\pi
 \frac{\cos2m\theta-\cos(2m-2)\theta}{1-2z\cos2\omega_{\theta}+z^2}\;d\theta,
\end{equation}
and we see (cf. \eqref{fnzeq2}) that we can express $G_m(z)$ in terms of two 
$F_m(z)$ functions. This gives equation \eqref{gmres}.
\end{proof}

\subsection{Lagrange inversion asymptotics}\label{lagrange}

Suppose we have an unknown function $w$ that we assume can be written as an 
(as yet unknown) power series.  All we know about this power series is that 
it can be written as a recursion relation 
\begin{align}\label{wdef}
  &w=z\varphi(w), &\varphi(0) \ne 0,
\end{align}
where $\varphi(w)$ is some generating function that is defined as an 
{\emph{implicit}} function of $z$.   We would like to find $w$ as 
an {\emph{explicit}} function of $z,$ so we can express some other 
function $f(w)$ as an explicit power series in $z.$  Lagrange Inversion 
enables us to do this, and tells us that we can write
\begin{equation}\label{genfun3}
 f(w)=\sum_{n \ge 1}\frac{t^{n}}{n}[\lambda^{n-1}]
                    f^{'}(\lambda)\phi^{n}(\lambda),
\end{equation}
where $\lambda$ is a dummy variable, $'$ denotes differentiation with 
respect to $\lambda$ and the square brackets $[x^n]$ is the 
``Goulden-Jackson'' notation \cite{Jackson} for the coefficient of the 
term in $x^n.$
\newline

Here are two very simple examples to illustrate the use of Lagrange 
Inversion.  Suppose we were given $w$ defined only to be a solution of 
the equation
\begin{equation}
 w=ze^{w},
\end{equation}
and we know nothing else about it.  Suppose we just want to obtain a power 
series for $f(w)=w,$ where $\varphi(w)=e^{w}.$ Then the formula above 
becomes
\begin{equation}\label{wsum1}
  w = \sum_{n \ge 1}\frac{t^{n}}{n}[\lambda^{n-1}]e^{n\lambda}
    =\sum_{n \ge 1}\frac{n^{n-1}}{n!}.
\end{equation}
It follows from Stirling's formula for the factorial that this series 
converges for $|z|<1/e$.
A slightly less simple example occurs if we are interested in the function 
$g(w) = w^2.$   Then equation \eqref{wsum1} would become 
\begin{equation}
 w^{2} = \sum_{n \ge 1}\frac{t^{n}}{n}[\lambda^{n-1}]2\lambda e^{n\lambda}
       = 2\sum_{n \ge 1}\frac{t^{n}}{n}[\lambda^{n-2}]e^{n\lambda}
       = 2 \sum_{n \ge 2} t^{n} \frac{n^{n-3}}{(n-2)!}.
\end{equation}

\subsubsection{Lagrange inversion for Jacobi polynomials}

The formula of Lagrange most useful for Jacobi polynomials is from
\cite{sriv} 
\begin{equation}\label{srivJ}
 \frac{f(\lambda)}{1-z \varphi'(\lambda)} = \sum_{n=0}^{\infty} 
            \frac{z^n}{n!}\frac{d^n}{dx^n} \{f(x) [\varphi(x)]^n \}.
\end{equation}
where $\varphi$ was defined in equation~\eqref{wdef} and $x$ is another 
dummy variable.

\bigskip

The asymptotics of the Jacobi polynomials have been discussed previously 
by Chen and Ismail \cite{ChenIsmail} and Ambainis {\emph{et al.}} used 
those results to derive some results on the asymptotics of the 
$\psi$-functions.  It should be noted that the Chen-Ismail results used the 
method of Darboux and are not uniform over the full range of $\alpha$.  
For the rest of this section we will briefly discuss Lagrange inversion 
and show how to derive the integral representations that Carteret 
{\emph{et al.}} used for the $\psi$-function to derive uniformly convergent 
asymptotics. 
Chen and Ismail's work on the Jacobi polynomials \cite{ChenIsmail} uses 
the generating function for $J_j^{(\gamma + aj, \beta + bj)}(0)$ of 
Srivastava and Singhal \cite{sriv} which in our notation becomes
\begin{equation}\label{genfun1}
 \sum_{j=0}^{\infty} J_{j}^{(\gamma + aj,\beta + bj)}(0)z^{j}
   =(1+{\bf{u}})^{\gamma+1}(1+{\bf{v}})^{\beta+1}[1-a{\bf{u}}-b{\bf{v}}
    -(1+a+b){\bf{uv}}]^{-1},
\end{equation}
where ${\bf{u}=-\bf{v}}$ and $\bf{u}$ is implicitly defined as a function 
of $z$ by 
\begin{equation}\label{genfunc2}
 -{\bf{u}}=\frac{z}{2}(1-{\bf{u}})^{1+a}(1+{\bf{u}})^{1+b},
\end{equation}
In order to obtain the asymptotics, we will need to interpret 
equation~\eqref{genfun1} in terms of equation~\eqref{srivJ}, again following 
Srivastava and Singhal \cite{sriv}.  
In the case of $\psi_L,$  we should let (see \cite{3routes}) 
\begin{align} 
 &a=\gamma=0, &\beta=\frac{1+\alpha}{1-\alpha}, 
 \quad \quad \quad \quad &b=\frac{2\alpha}{1-\alpha}.
\end{align}
Then we should define 
\begin{equation}
 \varphi(\lambda) = \frac{\lambda^2-1}{2}(1+\lambda)^{2\alpha/(1-\alpha)}
\end{equation}
and $f$ should be defined by equation~\eqref{srivJ}, which we must set 
equal to
\begin{equation}
 f(\lambda)=(1+\lambda)^{(1+\alpha)/(1-\alpha)},
\end{equation}
following the method originally developed in \cite{sriv}.  We omit the 
details, since we only wish to use the result.  Then
\begin{equation}
 J_{(t-n)/2-1}^{(0,n+1)}(0)=J_{m}^{(0+0\cdot m,\frac{1+\alpha}{1-\alpha}
+\frac{2\alpha m}{1-\alpha})}(0), 
\end{equation}
and so  
\begin{equation}
  m=(1-\alpha)t/2-1,
\end{equation}
as required. 

\bigskip

We would like to obtain an integral representation of the coefficients in
equation~\eqref{genfun1}.  
If $f(\lambda)$ and $\varphi(\lambda)$ are analytic then this can be done 
using the Cauchy integral formula, thus
\begin{equation}\label{genfun4}
[\lambda^{n}]f(\lambda)\phi^{n}(\lambda)
=\frac{1}{2 \pi i}\int_{C} f(\lambda) \phi^{n}(\lambda) 
\lambda^{-n-1}\;d\lambda
\end{equation}
where $C$ is a sufficiently small contour around the origin.  Equation
\eqref{genfun4} is an example of a Rodrigues formula for a set of orthogonal 
polynomials.  Another example of these appears in the method for generating 
orthogonal polynomials using Gram-Schmidt orthogonalization \cite{Arfken}.
So we obtain the integral representation for the Jacobi polynomials 
\begin{equation}
 J_{n}^{(0,2\alpha n/(1-\alpha)+\beta)}(0)
=\frac{1}{2\pi i}\int_{C}\frac{(1+\lambda)^{(1+\alpha)/(1-\alpha)}}{\lambda}
\left(\frac{\lambda^{2}-1}{2\lambda}
 (1+\lambda)^{2\alpha/(1-\alpha)}\right)^{n}
 \;d\lambda, 
\end{equation} 
as used in \cite{Saff,Gawyer}.
This is the contour integral that Saff and Varga estimated using steepest 
descents.  This is discussed in complete detail in \cite{3routes} so there 
is no need to say more here about this particular example. This example 
shows however that expressing a coefficient obtained using Lagrange 
inversion as a contour integral in this way and using steepest descents may 
lead to uniform asymptotic expansions over a wide domain. See the book 
\cite{AskeyRoy} by Andrews {\emph{et al.}} for examples of Lagrange 
inversion arising in special functions.

%----------------------%
%%% must end with... %%%
\end{document}